\documentclass[%
 reprint,
 amsmath,amssymb,
 aps,
]{revtex4-2}

\usepackage{graphicx}
\usepackage{dcolumn}
\usepackage{bm}

\usepackage{todonotes}

\DeclareMathOperator\erf{erf}

\begin{document}

\preprint{APS/123-QED}

\title{A Mean-Field Method for Generic Conductance-Based Integrate-and-Fire Neurons with Finite Timescales}

\author{Marcelo P. Becker}
 \email{marcelo.becker@bccn-berlin.de}
\altaffiliation[]{Bernstein Center for
Computational Neuroscience, 10115 Berlin, Germany}
\author{Marco A. P. Idiart}%
 \altaffiliation[]{Department of Physics, Institute of Physics, Federal University of Rio Grande do Sul, Porto Alegre, Brazil}




\date{\today}

\begin{abstract}
The construction of transfer functions in theoretical neuroscience plays an important role in determining the spiking rate behavior of neurons in networks. These functions can be obtained through various fitting methods, but the biological relevance of the parameters is not always clear. However, for stationary inputs, such functions can be obtained without the adjustment of free parameters by using mean-field methods. In this work, we expand current Fokker-Planck approaches to account for the concurrent influence of colored and multiplicative noise terms on generic conductance-based integrate-and-fire neurons. We reduce the resulting stochastic system from the application of the diffusion approximation to a one-dimensional Langevin equation. An effective Fokker-Planck is then constructed using Fox Theory, which is solved numerically to obtain the transfer function. The solution is capable of reproducing the transfer function behavior of simulated neurons across a wide range of parameters. The method can also be easily extended to account for different sources of noise with various multiplicative terms, and it can be used in other types of problems in principle.

\end{abstract}

\maketitle


\section{\label{sec:level1}Introduction}

The brain is a complex system that organizes itself into structures ranging in size from fine-scale components \cite{Yoshimura2005} to large-scale arrangements involving the whole organ \cite{Koziol2014-pi}. These structures are composed of neurons and supportive cells that interact with each other in complex ways \cite{Vogels2005-xq}, presenting a typical phenomenon of complex systems. As a result, the study of these systems poses significant challenges, requiring the development of theoretical models and frameworks to bridge the multiple scales.

Theoretical models and frameworks are essential to address the multi-scale challenges found in complex systems. Nature has many examples of such systems, and condensed matter physics has developed a variety of methods to treat these types of problems \cite{Stanley1971-ft, mandelbrot1984, chen2014, de2016}. These methods have been co-opted by theoretical neuroscientists to understand the complex interactions of neurons and supportive cells in the brain \cite{Helias2019-wz}.

Mean-field methods are an effective tool for investigating collective behaviors in neural networks, as neurons are often influenced by numerous stochastic inputs. By modeling neurons as simple transfer functions, attractor dynamics, network oscillations, synchronization, pattern formation, and phase transitions can be better analyzed \cite{Bienenstock1982-ek,Ermentrout1998-le,Nicolas_Brunel1999-yr,Brunel2000-uu}. Through mean-field approaches, a connection can be made between the individual spiking neuron at the microscopic scale and the population's rate descriptions at the mesoscopic scale \cite{Brunel2000-uu, Schwalger2017-vv}. Amit and Brunel's seminal work introduced a typical approach using the Fokker-Planck mean-field formalism to model the simple leaky integrate-and-fire model \cite{Amit1997-hp}.

However, the mean-field treatment of conductance-based integrate-and-fire neurons introduces a new level of complexity due to the conductance variables, which increase the dimensionality of the stochastic system. These complexities can generate measurable differences, such as the temporal correlations introduced by synaptic filtering, which have been shown to alter the spiking statistics \cite{Lindner2004-zj} and the scaling properties of neurons \cite{Sanzeni2020-ab}. Dealing with these temporal correlations presents a challenge for mean-field methods. Perturbative solutions have been found for a single source of additive exponentially correlated noise \cite{Brunel1998-hw, Moreno-Bote2004-bk}. For linear multiplicative noise, it is possible to use the same perturbative methods by employing the effective time-constant approximation \cite{Richardson2004-qu, Richardson2005-it}. By combining these two methods, a large set of problems with linear multiplicative terms can be treated. But what happens if we want to expand those methods to include more generic forms of neurons that include colored noise with nonlinear multiplicative terms?

Looking back at physics, the presence of temporal correlations in colored noise processes is known to prevent the construction of an exact Fokker-Planck equation. Some forms of approximation were derived in the 80s. The best Fokker-Planck approximation, proposed by Sancho and Lindenberg \cite{Sancho1982-ho, Lindenberg1983-pv}, proposes a differential form for the diffusion term that is exact in the white noise limit but only solvable in particular cases. Therefore, perturbative methods are required in most applications. The Fox theory, which uses functional calculus to derive an approximation for small correlations \cite{Fox1986-gx, Fox1986-uq}, yields the same result as the best Fokker-Planck approximation at first order in the time correlation but diverges in higher orders \cite{Grigolini1988-eg}. Other methods, such as the projection operator method \cite{Faetti1987-mr, Grigolini1988-eg}, the adiabatic elimination procedure used by Jung and Hänggi \cite{Jung1987-md}, and the renormalized operator cumulant expansion of Der \cite{Der1989-sy}, have also been developed to deal with similar problems. 

In this work, we adapted the Fox Theory to the case of a conductance-based integrate-and-fire neuron under the influence of stochastic inputs. We numerically solved the resulting stationary Fokker-Planck equation and extracted the transfer function with different assumptions on the boundary conditions, comparing the results with proper simulations.

\section{General model}
We consider the behavior of a point-like generic leaky integrate-and-fire neuron with conductance-based input embedded in a network of similar units. The neuron is described by the membrane potential $V$ that follows from
\begin{equation}\label{Veq}
    \tau_L\frac{dV}{dt} = -(V - E_L) - \sum_i g_i(t)s_i(V)(V - E_i)\,,
\end{equation}
where $\tau_L$ is the membrane time constant, $E_L$ is the resting potential, $E_i$ is the reversal potential of the corresponding channel $i$, and $s_i(V)$ is a modulating function that can depend on $V$. It is important to note that the addition of nonlinear functions of $V$ to the equation (a quadratic or exponential function for example) is possible in principle, although we will not deal with this case here. The conductances $g_i(t)$ behave as linear filters of the input signal. Specifically, we have
\begin{equation}\label{dgE}
    \tau_i \frac{dg_i}{dt} = -g_i + w_i \sum _{j,k}\delta(t - t_j^k)\,.
\end{equation}
The summation here is performed over all pre-synaptic sites $j$ and all spikes $k$ emitted in that site. $w_i$'s are the synaptic weights, which are kept the same for all neurons belonging to the same population.

The membrane potential $V$ evolves according to \eqref{Veq} until it reaches the threshold $\theta$ when a formal spike is emitted. The potential is then reset to $V_r$ and is not updated for the extent of the refractory interval $\tau_{r}$.

\section{Mean-field analysis}
\subsection{Conductance}
As a starting point, we suppose that the neuron receives inputs from separate populations of neurons corresponding to different channels in the equation, each of them making $K_i$ connections. We assume that the inputs from each population come from Poisson rate neurons with fixed rate $\nu_i$. If the number of connections is large  ($K_i \gg 1$) and the connection weights small ($w_i \ll 1$) the diffusion approximation can be used \cite{Amit1997-hp, Feng2003-hw} and equation \eqref{dgE} becomes

\begin{equation}\label{difgE}
    \tau_i \frac{dg_i}{dt} = -g_i + \mu_i + \sqrt{\tau_i}\sigma_i \xi_i(t)\,,
\end{equation}
with
\begin{eqnarray}
    \mu_i = w_iK_i\nu \tau_i\,,
\\
    \sigma_i^2 = w_i^2K_i\nu \tau_i\,.
\end{eqnarray}
The $\xi_i(t)$'s here are uncorrelated Gaussian variables with zero mean and unit variance.

\subsection{Membrane Potential}
The membrane potential equation \eqref{Veq} can then be written as
\begin{equation}\label{LGV}
    \frac{dV}{dt} = -\frac{(V - \mu)}{\tau} + \sum_i h_i(V) \eta_i(t)\,,
\end{equation}
where 
\begin{eqnarray}
    &\tau = \tau_L/(1 + \sum_i s_i(V)\mu_i )\,,
\\ \nonumber \\
    &\mu = \frac{\tau}{\tau_L}(E_L + \sum_i s_i(V)\mu_i E_i)\,,
\\ \nonumber \\
    &h_i(V) = s_i(V)\frac{\sqrt{\tau_i}}{\tau_L}\sigma_i(E_i - V)\,,
\end{eqnarray}
and the noise variables are now 
\begin{equation}
    \eta_i(t) = \frac{1}{\tau_{i}}\int_0^t e^{-\frac{T}{\tau_i}}\xi_i(t-T)dT\,,
\end{equation}
with the following correlations
\begin{equation}
    \langle \; \eta_i(t) \; \eta_j(t') \; \rangle = \frac{1}{2\tau_i}e^{-\frac{|t-t'|}{\tau_i}} \delta_{ij} \;\;.
\end{equation}

What we have now is a Langevin equation with distinct sources of colored noise. The $n$ dimensional stochastic system is now reduced to a single SDE. This happens, however, with the cost that the noise is no longer Markovian, i.e., the fluctuations at time $t$ depend on the fluctuations at previous times $t' < t$. It makes it impossible to obtain an exact Fokker-Planck equation since it adds an additional temporal integration.  Therefore, in order to be able to use the Fokker-Planck approach \cite{Amit1997-hp}, it is necessary to build an approximate Fokker-Planck equation. 

\subsection{Effective Time-Constant Approximation}
In some cases, it might also be useful to simplify the problem by finding an approximation that allows us to eliminate the multiplicative noise. Observe that in Eq. \eqref{LGV} the stochastic variable $\eta_i(t)$ is multiplied by a function of $V$  implying that the noise level depends on voltage values.  This complexity can be avoided by using the effective time-constant approximation \cite{Richardson2004-qu}\cite{Richardson2005-it} where the membrane potential is replaced by the equilibrium potential $\mu$, resulting in
\begin{equation*}
    h_i(V) \rightarrow h_i(\mu)
\end{equation*}
This approximation implies that the modulation of noise can be seen as dependent on the distance of the equilibrium potential from its reversal potential, at least at first order. In fact, it can be argued that using this approximation, when the terms $h_i(V)$ are linear, leads to a more consistent treatment of the problem, since the error generated by this approach is of the same order as the error introduced by the diffusion approximation \cite{Richardson2005-it}. But the most important fact here is that this approach simplifies considerably the treatment of the resulting Fokker-Planck equation. A direct consequence of this is that it lends to more easily interpretable parameters. We will compare the results with and without the use of this approximation when it is applicable.

\subsection{Fox Theory}
Temporal correlations in the noise of Langevin equations are known to impede the construction of an exact Fokker-Planck equation. So our task is to find an appropriate approximation that leads to a differential equation of the probability distribution of the membrane potential. From the different options available, we use here the Fox theory, since it is the one with the most direct application and is the easiest to generalize for multiple noise sources. It also possesses some relevant properties for this work. First of all, it is important to note that the approach followed in this method is non-perturbative. In fact, under a certain condition, the convergence of the approximation for $\tau_i \rightarrow 0$ is uniform, that is, it converges to the white noise case for all values of $V$ in the domain of interest \cite{Fox1986-uq}. The uniformity condition is
\begin{equation}\label{FT:UCcondition}
    1 - \tau_i\left( W'(V) - \frac{h_i'(V)}{h_i(V)}W(V)\right) > 0\,,
\end{equation}
where $W(V)$ is the drift term in the Langevin (in our case $(\mu - V)/\tau$) and primes indicate derivatives with respect to the argument $V$. This condition sets a scale for $\tau_i$ for which the approximation behaves reasonably well.

Jung and Hänggi adiabatic method \cite{Jung1987-md} is an approximation that is valid for small and large values $\tau_i$, whose stationary solution agrees with the stationary solution of the Fox theory. Therefore, even though the resulting effective Fokker-Planck obtained by the Fox theory was derived for small $\tau_i$ values, the stationary solution is valid also for large $\tau_i$ (given that condition \eqref{FT:UCcondition} is obeyed). The validity for both limits suggests that the Fox theory is a good interpolation between both stationary results. A small derivation of the validity of the stationary solution of the Fox theory for $\tau_i \rightarrow \infty$ can also be found in \cite{Grigolini1988-eg}.

The application of the Fox Theory to equation \eqref{LGV} results in the effective Fokker-Planck
\begin{equation}\label{FP}
   \frac{\partial P}{\partial t} = -\frac{\partial}{\partial V}\left[W(V)P - \sum_i h_i(V) \frac{\partial}{\partial V} (S_i(V)P)\right]
\end{equation}
with
\begin{equation}\label{eq:MN_S}
    S_i(V) = \frac{1}{2}\left[ \frac{h_i(V)}{1 - \tau_i(W'(V) - \frac{h_i'(V)}{h_i(V)}W(V))}\right]\,
\end{equation}
The expansion for multiple noise sources can be simply done by using the same assumptions as in the original papers \cite{Fox1986-gx,Fox1986-uq}.

\subsection{Transfer Function}
First, we will explore the result of the application of the effective time-constant approximation. The resulting effective Fokker-Planck is then
\begin{equation}\label{TF:FP}
    \frac{\partial P(V, t)}{\partial t} = \frac{\partial}{\partial V}\left[ \frac{(V - \mu)}{\tau}P(V,t)\right] + \frac{\sigma_V^2}{2\tau}\frac{\partial^2 P(V,t)}{\partial V^2}\,,
\end{equation}
where 
\begin{equation}\label{TF:sigma}
    \sigma_V^2 = \sum_i \sigma_{V_i}^2 =  \sum_i\frac{\tau^2}{\tau + \tau_i}h_i^2\,.
\end{equation}
and the notation was simplified by calling $h_i(\mu ) = h_i$.

The resulting effective Fokker-Planck is the same as the one obtained from the simpler Langevin
\begin{equation}
    \frac{dV}{dt} = -\frac{(V-\mu)}{\tau} + \sigma_V \; \xi(t)
\end{equation}
with the zero mean unit variance Gaussian white noise $\xi(t)$.

The independence of the terms in the sum suggests the separation of the full variance into two independent components. There is, then, a simple interpretation of the parameters of the effective Fokker-Planck equation. The first-order term (drift) corresponds to the deterministic drive of the membrane potential. The second-order term (diffusion) can be seen as the sum of the variance of the noise sources, where the noise sources are treated as white. The combination of both approximations (effective time-constant and the Fox theory), therefore, produces the same behavior as a Langevin system with white noise and summed variances.

The transfer function can be calculated, resulting in
\begin{equation}\label{TF:lTF}
    \frac{1}{\nu} = \tau_r + \tau\sqrt{\pi} \int^{\frac{\theta - \mu}{\sigma_V}}_{\frac{V_r - \mu}{\sigma_V}}e^{x^2}(1 + \erf(x))dx\,,
\end{equation}
where $\erf(x)$ is the error function. We can also get the stationary probability distribution
\begin{eqnarray}\label{TF:PD}
    P_S(V) &=& \frac{2\nu \tau}{\sigma_V}\exp\left(-\frac{(V - \mu)^2}{\sigma_V^2}\right) \nonumber \\
    & & \times \int^{\frac{\theta - \mu}{\sigma_V}}_{\frac{V - \mu}{\sigma_V}}\Theta\left(x - \frac{V_r - \mu}{\sigma_V}\right)e^{x^2}dx\,,
\end{eqnarray}
where $\Theta(x)$ is the Heaviside function.
In these results, there is an implicit assumption of the continuity of the distribution at the threshold, which implies $P_S(\theta)=0$. This assumption is not reasonable, since perturbative results for single channel conductance-based models show a discontinuity at that point \cite{Brunel1998-hw}. We can estimate numerically the value of the distribution in a procedure that will be described shortly. We will provide, also, comparison between results with and without the assumption.

\subsection{Multiplicative Noise}
The complicated form of \eqref{FP} allows only a formal stationary solution. Using again the continuity of the distribution and standard procedures \cite{Feng2003-hw}, we arrive in
\begin{equation}
    \frac{1}{\nu} = \tau_r + \int_{V_r}^\theta \int_{-\infty}^x \frac{e^{F(x) - F(V)}}{\chi(x)} \;dVdx\,,
\end{equation}
where
\begin{eqnarray}
    F(V) = \int \frac{\sum_ih_i(V)\frac{dS_i}{dV}  - W(V)}{\chi(V)} \; dV
\end{eqnarray} 
and
\begin{eqnarray}
    \chi(V) = \sum_i h_i(V) S_i(V)\,.
\end{eqnarray}
The lower limit of the first interval can be limited by the lowest of the reversal potentials since the dynamics of the membrane potential cannot be lower than this value. But in a general sense, the infinity can be written in place as we did.

This formal solution, unfortunately, is not straightforward to use, since a closed form for the integrating factor is generally not obtainable. Therefore, we have to rely on numerical methods.

\subsection{Numerical Methods and Simulations}
Efficient results can be obtained by using the numerical approach developed by Richardson \cite{Richardson2007-kf}. It takes advantage of the formal solution to get faster convergence than a typical Euler integration and uses $P_S(\theta)=0$ as an initial condition for the integration. However, since this assumption is not well founded in our case, we need to develop a different approach. We opted for using a double integration procedure, starting with the more reasonable assumption $P(E_I) = 0$ and integrating forward. This results in an estimated value for $P_S(\theta)$, so that we can backward integrate to obtain the distribution and the firing rate.

The simulation data was generated using Brian2 \cite{Stimberg2019} treating the input layer as Poisson neurons. Firing rates were taken as the time average of the spikes for a period of 10s after waiting for 5s to eliminate transients. The same strategy was used for the generation of the distribution.

\section{Conductance-Based Integrate-and-Fire Neuron}
We will first apply our method to a simple conductance-based integrate-and-fire neuron with two input channels: one excitatory $g_E(t)$, and one inhibitory $g_I(t)$. The equations describing this system are
\begin{eqnarray}
    &\tau_L \frac{dV}{dt} = -(V - E_L) - \sum_{i=E,I}g_i(t)(V - E_i)\,,\\
    &\tau_E \frac{dg_E}{dt} = -g_E + \sum_{j,k} w_E \delta(t - t_j^k)\,,\\
    &\tau_I \frac{dg_I}{dt} = -g_I + \sum_{j,k} w_I \delta(t - t_j^k)\,.
\end{eqnarray}
The neuron receives input from $K_E$ excitatory input neurons and $K_I$ inhibitory. Both populations fire at a fixed firing rate $\nu_i$. The parameter values are chosen to be physiologically plausible and are in the range typically used in simulation works (for instance \cite{Zenke2014-im}), see table \ref{tab:CoBaIF}.

\begin{ruledtabular}\textbf{}
\begin{table}
    \centering
    \begin{tabular}{c c}
    \textbf{Parameter} & \textbf{Value} \\ \colrule 
    $E_L$ & -60mV  \\
    $E_E$ & 0mV  \\
    $E_I$ & -80mV  \\
    $w_E$ & \{0.1, 0.5\}  \\
    $w_I$ & \{0.1, 0.4, 1.0, 10.0\}  \\
    $\tau_L$ & 20ms  \\
    $\tau_E$ & variable  \\
    $\tau_I$ & 10ms  \\
    $\tau_R$ & 2ms  \\
    $K_E$ & 400  \\
    $K_I$ & 100  \\
    $\theta$ & -50mV  \\
    $V_r$ & -60mV \\
    $\nu_i$ & \{5, 20, 50\}Hz
    \end{tabular}
    \caption{Table containing the parameters used for the simple Conductance-Based Integrate-and-Fire model.}
    \label{tab:CoBaIF}
\end{table}
\end{ruledtabular}

\begin{figure}
    \centering
    \includegraphics[width=0.48\textwidth]{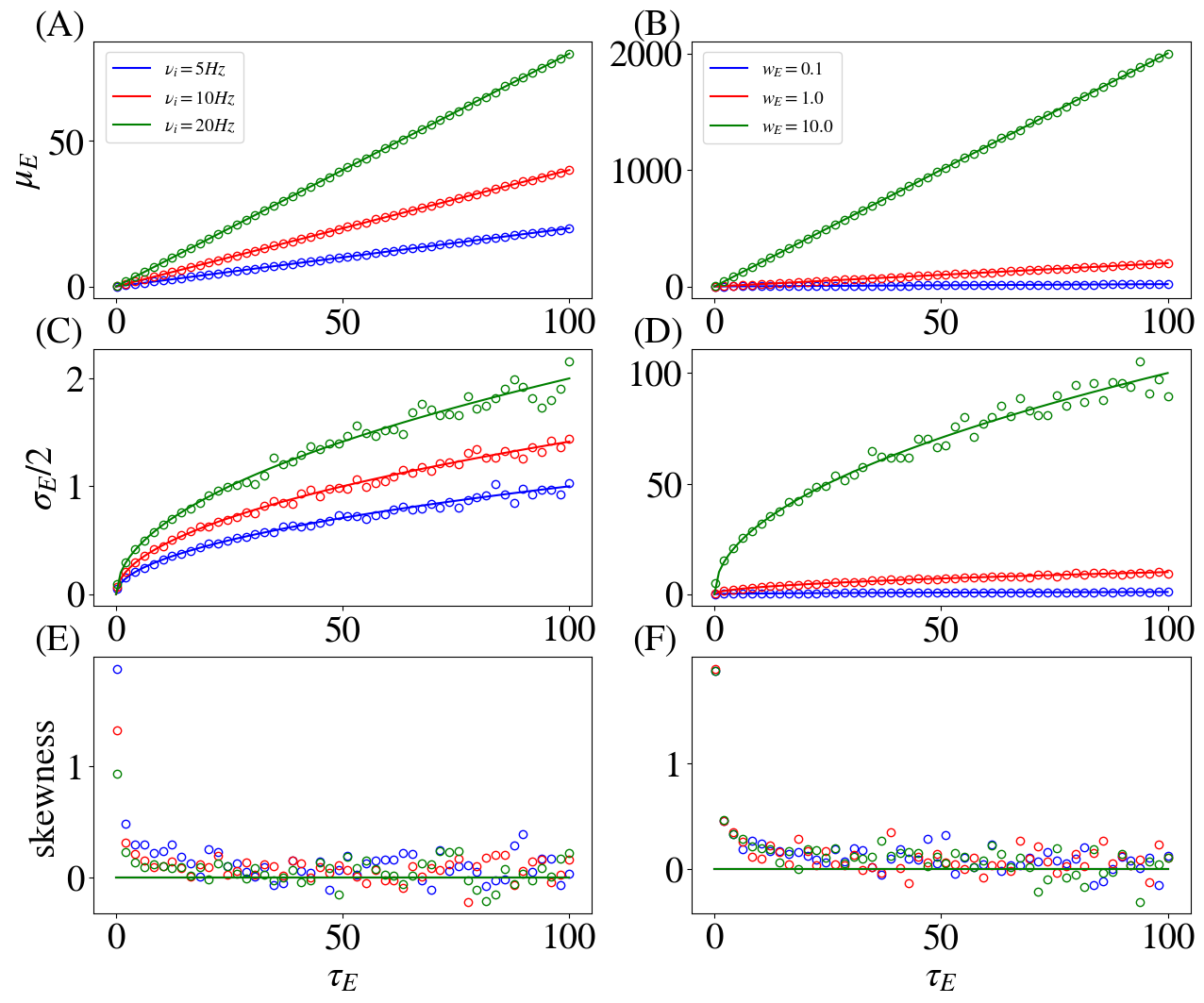}
    \caption{Comparison of the analytical expressions obtained for the statistics of $g_E$ using the diffusion approximation (lines) with simulations (circles). The analytical expressions are good descriptions of the simulations for all the range of parameters tested for the first and second moments. The skewness, however, exhibits a deviation from the Gaussian approximation for small $\tau_E$. Parameters for first column, $w_E = 0.1$, $w_I = 0.4$; second column, $w_I = 0.8$, $\nu_i = 5Hz$}
    \label{fig:g_comp}
\end{figure}

To illustrate the validity of the diffusion approximation, for the range of values used here, we plotted the mean, the standard deviation, and the skewness of the analytical conductance and the simulated one (figure \ref{fig:g_comp}). As expected by the construction, the expressions for the mean and the standard deviation are a good representation of the values simulated even when the synaptic weight $w_E$ is large. In contrast, we can see deviations in the skewness for small values of $\tau_E$. This is somewhat expected since the values of $g(t)$ can't be negative and the diffusion approximation doesn't take this into consideration. For small $\tau_E$, we have small $\mu_E$ and the Gaussian form of the approximation fails to account for the asymmetric shape of the distribution with a hard boundary at $g=0$. Therefore, as stated in \cite{Richardson2005-it}, the diffusion approximation introduces errors at the third-order moment of the distribution.

The resulting form of the Langevin equation for the Conductance-Based Integrate-and-Fire is
\begin{equation}
    \frac{dV}{dt} = -\frac{(V - \mu)}{\tau} + h_E(V)\; \eta_E(t) + h_I(V)\; \eta_I(t)\,,
\end{equation}
where
\begin{eqnarray*}
    &\tau = \frac{\tau_L}{1 + \mu_E + \mu_I}\,, \\
    &\mu = \frac{\tau}{\tau_L}(E_L + \mu_E E_E + \mu_I E_I)\,, \\
    &h_{E,I}(V) = \frac{\sqrt{\tau_{E,I}}}{\tau_L}\sigma_{E,I}(E_{E,I} - V)\,.
\end{eqnarray*}

We can now use the effective time-constant approximation or deal with the full multiplicative problem. 

\subsection{Additive Noise}
\begin{figure*}
    \centering
    \includegraphics[width=0.99\textwidth]{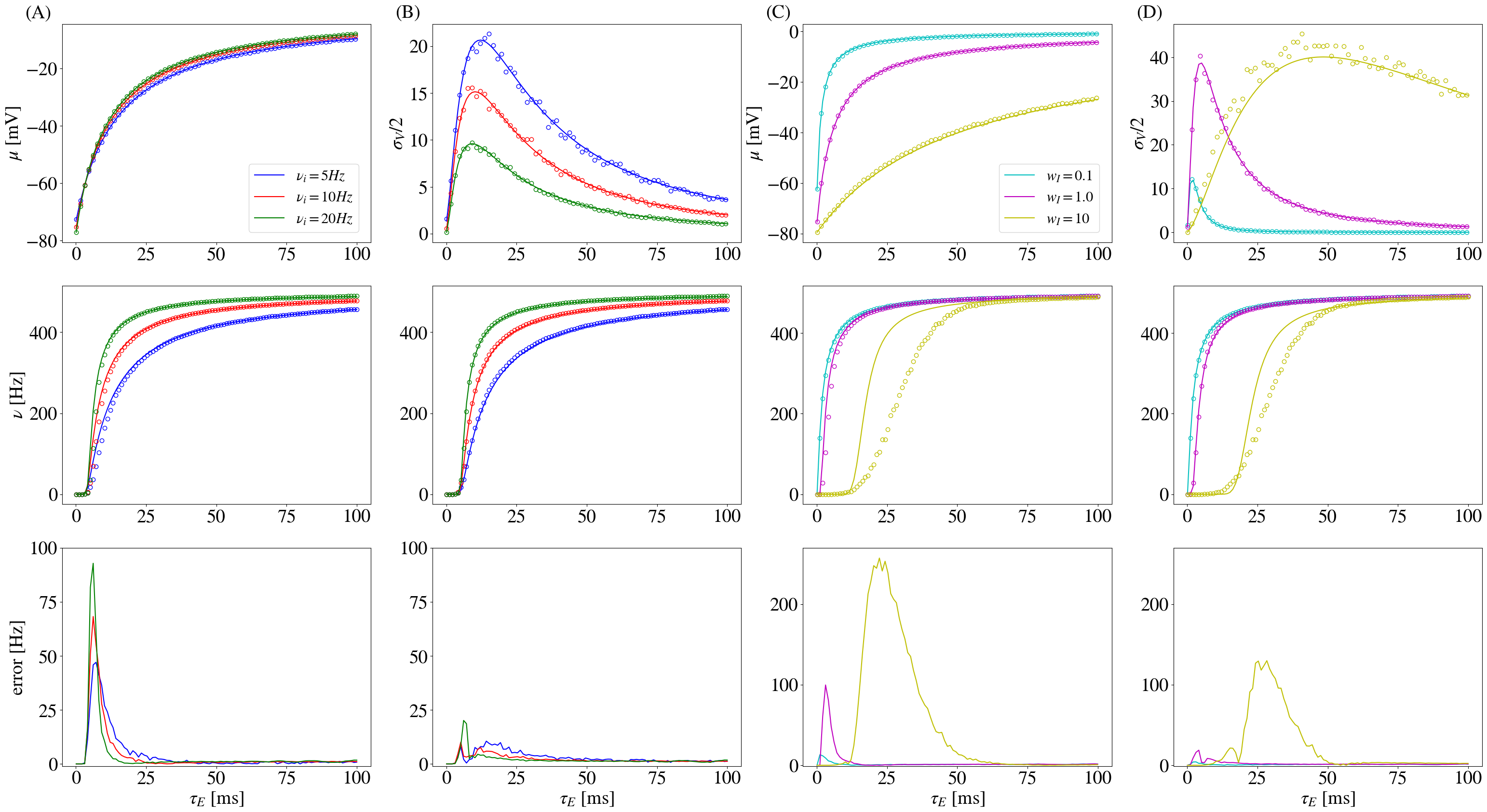}
    \caption{Comparison of the analytical model (lines) with simulations (circles)  for the simple conductance-based integrate-and-fire neuron using the effective time-constant approximation. In columns (A) and (B), we use three different values of input firing rate $\nu_i$ as a function of the excitatory time constant and set $w_E=0.1$ and $w_I=0.4$. In columns (C) and (D), we compare three values of inhibitory weights $w_I$ and set $w_E=0.5$, $\nu_i=5Hz$. In the first row and columns (A) and (C) the mean potential for a thresholdless model is plotted. Columns (B) and (D) show the standard deviation of the membrane potential for the same thresholdless model. In the second row,  columns (A) and (C) display the firing rates for the model with the continuous distribution assumption, and columns (B) and (D) display the firing rates using the double integration procedure. The absolute errors for the corresponding models are plotted in the third row.}
    \label{fig:AN_irate_comp}
\end{figure*}

With the effective time-constant approximation, the Langevin simplifies to
\begin{equation}
    \frac{dV}{dt} = -\frac{(V - \mu)}{\tau} + h_E\eta_E(t) + h_I\eta_I(t)\,,
\end{equation}
where the constant coefficients are $h_E = h_E(\mu)$ and $h_I = h_I(\mu)$. The application of the Fox Theory results in the transfer function \eqref{TF:lTF} and the probability distribution \eqref{TF:PD} with the expression for $\sigma_V$ given by
\begin{equation}
    \sigma_V^2 = \frac{\tau^2}{\tau + \tau_E}h_E^2 + \frac{\tau^2}{\tau + \tau_I}h_I^2\,.
\end{equation} A comparison of the analytical results (calculated numerically) with the simulations can be seen in Fig.\ref{fig:AN_irate_comp}. Fig.\ref{fig:AN_irate_comp}A and Fig.\ref{fig:AN_irate_comp}C assume $P(\theta) = 0$, while Fig.\ref{fig:AN_irate_comp}B and Fig.\ref{fig:AN_irate_comp}D do not. Good agreement is present for most sets of parameters tested, the exception being the high inhibition regime ($w_I=10$). Fig.\ref{fig:AN_irate_comp}A and \ref{fig:AN_irate_comp}B, show that the input rate $\nu_i$ has little effect on the stationary potential $\mu$ but changes the noise variance $\sigma_V^2$ of the neuron without threshold. Therefore, in this case,  the firing rate behavior is mostly given by the changes in the input variance. A higher noise variance makes the transition from silent to firing smoother, as can be seen by the slower convergence of the firing rate curves with higher input variance.

\noindent
Figs.\ref{fig:AN_irate_comp}C and \ref{fig:AN_irate_comp}D explore the effect of varying the inhibitory synaptic weight ( $w_I = 0.1$, $w_I = 1.0$, and $w_I = 10.0$) for a constant excitatory weight $w_E=0.5$. For low inhibition, there is a good agreement between theory and simulation. But for high inhibition ( $w_I=10.0$) the theory produces a sharp transition of firing rate that is not observed in the simulations. The discrepancy (see the error in Fig.\ref{fig:AN_irate_comp}C, third row) is larger in the region where $\mu$ is between $E_L$ and $\theta$, that is, in the sub-threshold regime. In this region, spikes are driven by membrane potential fluctuations. The sharpness of the transition compared to the data suggests that, for those parameter values, the model underestimates fluctuations. The high inhibition case also allows us to see that the $P(\theta) = 0$ brings the transition to lower $\tau_E$ values. The better result of the double integration procedure stems from the improvement of the estimation of the transition region since the shape of the curve is almost the same.

\subsection{Multiplicative Noise}
The full-multiplicative noise treatment results in a Fokker-Planck equation with the form
\begin{equation}
    \frac{\partial P}{\partial t} = -\frac{\partial}{\partial V}\left[ W(V)P \right] + \sum_{i=E,I}\frac{\partial}{\partial V}h_i(V)\frac{\partial}{\partial V}\left( S_i(V)P\right),
\end{equation}
where the functions $S_E(V)$ and $S_I(V)$ are given by the generic expression in \eqref{eq:MN_S}. The stationary differential equation that needs to be solved numerically is then
\begin{equation}
    \frac{\partial P_s}{\partial V} + B(V)P_s = -\nu H(V)\,,
\end{equation}
with
\begin{eqnarray}
    &B(V) = \frac{h_E(V)S'_E(V) + h_I(V)S'_I(V) - W(V)}{\chi(V)}\,,\\
    &H(V) = \frac{\Theta(V - V_r)}{\chi(V)}\,,\\
    &\chi(V) = h_E(V)S_E(V) + h_I(V)S_I(V)\,.
\end{eqnarray}

\begin{figure*}
    \centering
    \includegraphics[width=0.99\textwidth]{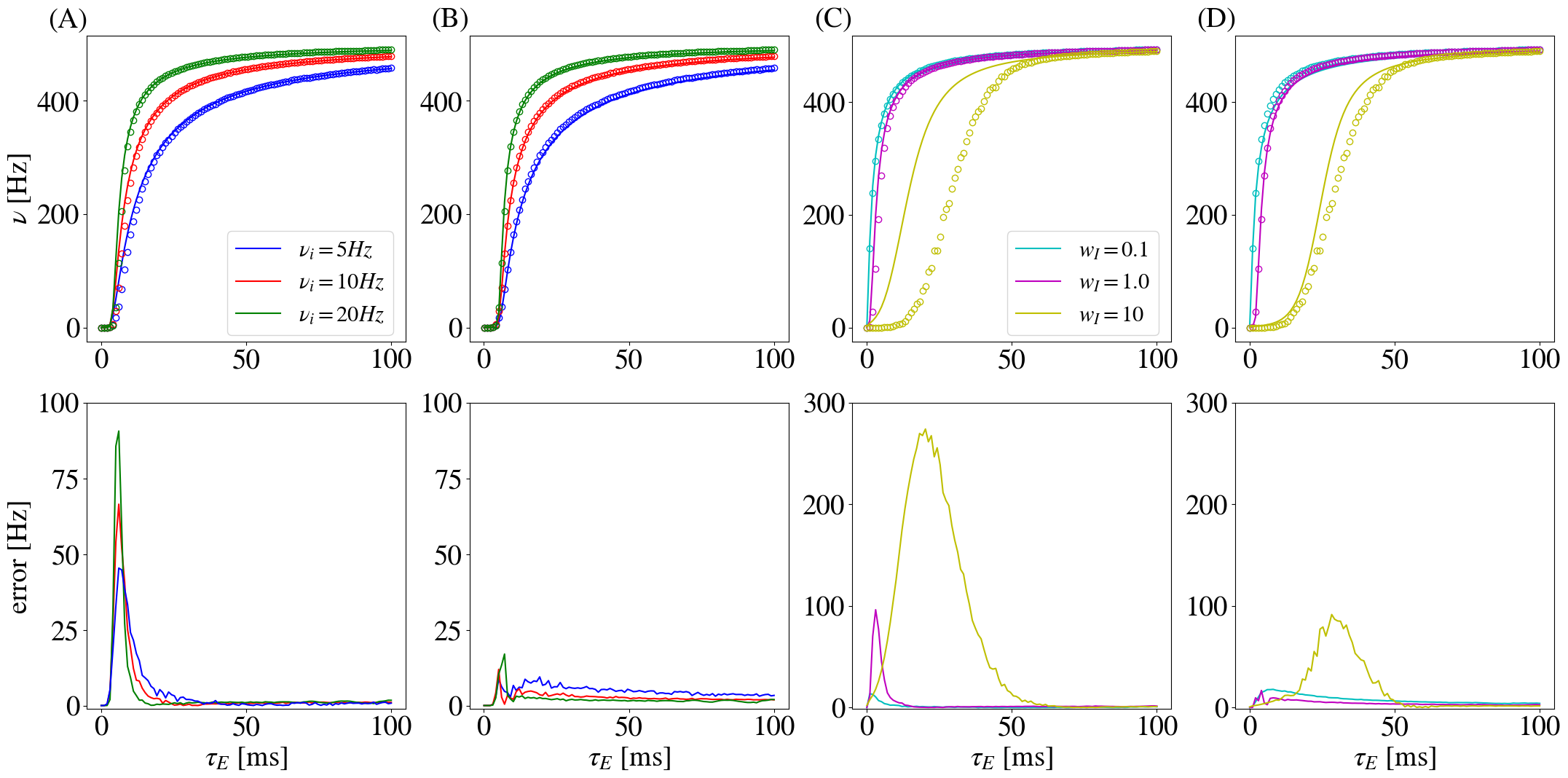}
    \caption{We compare the analytical model with simulations for the simple conductance-based integrate-and-fire neuron, incorporating full multiplicative noise. In columns (A) and (B), we vary the input firing rate $\nu_i$ as a function of the excitatory time constant and set $w_E=0.1$ and $w_I=0.4$. In columns (C) and (D), we compare different values of inhibitory weight $w_I$ and set $w_E=0.5$, $\nu=5Hz$. In the first row, columns (A) and (C) display the firing rate for the model assuming continuous distribution, and columns (B) and (D) display the firing rates using the double integration procedure. The absolute errors for the corresponding models are plotted on the second row.}
    \label{fig:MN_frate_mult}
\end{figure*}

We used this approach to solve the stationary Fokker-Planck equation numerically for the same set of parameters as in the last subsection.No appreciable differences were found between the additive and multiplicative models for different input firing rates (see Figs. \ref{fig:MN_frate_mult}A and \ref{fig:MN_frate_mult}C). However, for high inhibition ($w_I=10.0$), the full multiplicative treatment with the continuity assumption ($P(\theta)=0$) produced a worse quantitative result but with a better overall shape of the curve. This error was mostly corrected when we dropped this assumption and used the double integration procedure (see Fig. \ref{fig:MN_frate_mult}D), concentrating on the region where $\mu$ is around the threshold value.

\subsection{Stationary Probability Distribution}
\begin{figure*}
    \centering
    \includegraphics[width=0.99\textwidth]{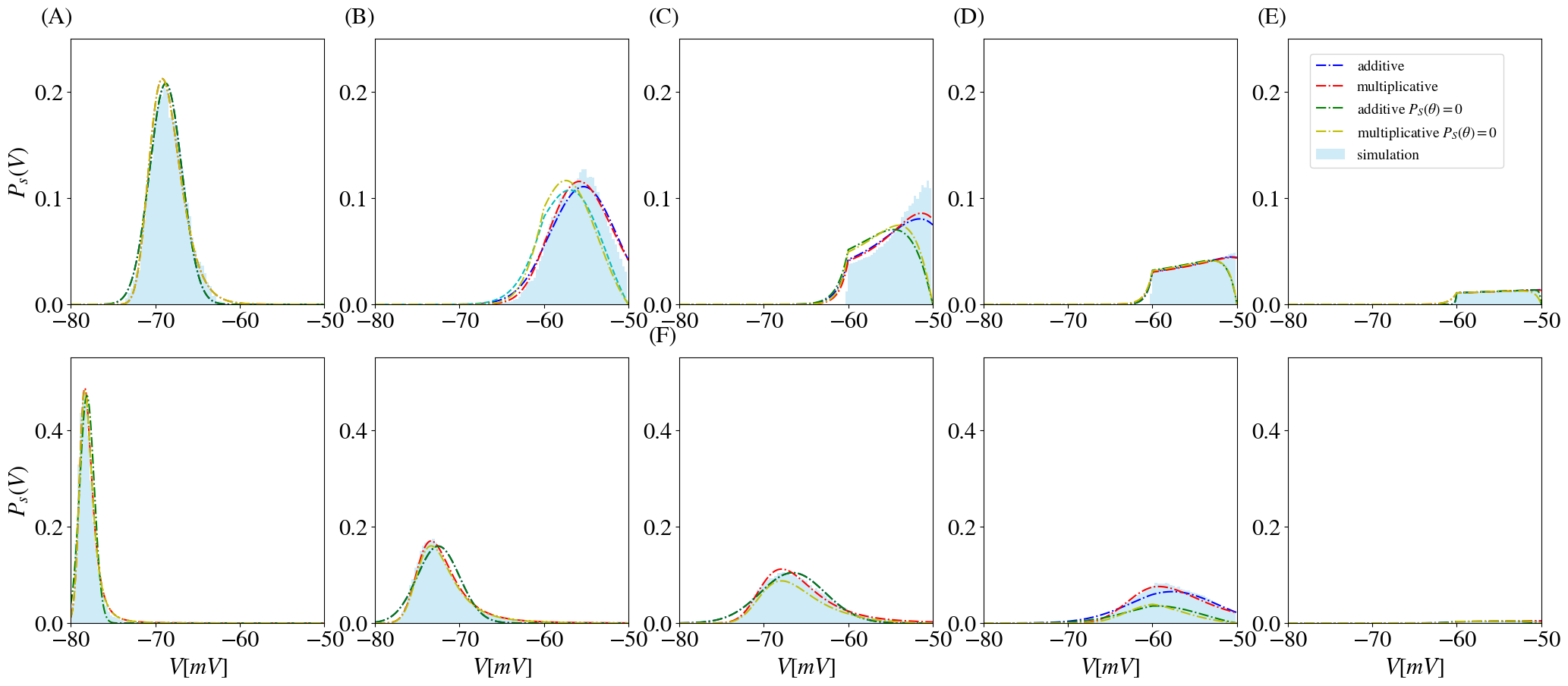}
    \caption{ The stationary probability distributions for the Voltage of conductance-based integrate-and-fire neurons. Column (A) $\tau_E = 1ms$, (B) $\tau_E = 5ms$, (C) $\tau_E = 10ms$, (D) $\tau_E = 20ms$, (E) $\tau_E = 70ms$. The first row corresponds to blue curves in Fig. \ref{fig:AN_irate_comp}, that is, $\nu_i=5Hz$ and parameters from table \ref{tab:CoBaIF}. The second row corresponds to the yellow curves on the same figure ($w_I = 10$, $w_E = 5$, and $\nu_i = 5Hz$). The different lines correspond to different assumptions in the model, as can be seen in the legend in column (E).}
    \label{fig:MN_dist}
\end{figure*}

To complete the analysis of the conductance-based integrate-and-fire neuron, we look at the stationary probability distributions with and without the effective time-constant approximation and the $P_S(\theta) = 0$ assumption. Fig.\ref{fig:MN_dist} top row corresponds to the parameters of the blue curves that appear in Figs.\ref{fig:AN_irate_comp} and \ref{fig:MN_frate_mult}.  Fig.\ref{fig:MN_dist} bottom row corresponds to the yellow curves of the same figures. Each column corresponds to progressive values of $\tau_E$ (1, 5, 10, 20, 70\;ms, respectively). As the excitatory time constant increases, the voltage distribution evolves from a Gaussian shape with a sharp peak far from the threshold, towards an increasingly more distorted distribution as it interacts with the threshold. The area under the curve diminishes (since we are omitting the refractory period in the graph) corresponding to the higher firing rate of the neuron.

Comparing the mean-field solutions, we can see that the major difference comes from the continuity assumption for intermediary values of $\tau_E$. It is clear that the simulated distributions are not continuous. In fact, when in the mean-driven regime ($\mu > \theta$), the distributions concentrate between the reset and the threshold. The double integration procedure gives good results for the distribution in most cases, with an artificial low-end tail for some values. The difference generated by the effective time-constant approximation is minor, generating almost superimposed curves in some cases.

\section{NMDA Integrate-and-Fire}
In the standard conductance-based integrate-and-fire neuron model, there is no $V$ nonlinearity in the resulting Langevin equation. In principle, our mean-field method should be able to handle nonlinearities in either the drift term, the diffusion terms, or both. Here we will introduce a nonlinearity by adding NMDA channels, which are excitatory channels whose activation depends on the membrane potential. A convenient way to model its behavior is by adding an appropriately tuned sigmoidal factor to the conductance term \cite{Jahr1990-ga}. The complete model can be written as
\begin{eqnarray}\label{eq:NMDA-model}
    \tau_L \frac{d V}{dt} = -(V - E_L) - (1 - \alpha) g_A(t)(V - E_E) - \nonumber \\
    - \alpha s(V)g_N(t)(V - E_E) - g_I(t)(V - E_I)\,,
\end{eqnarray}
where
\begin{eqnarray}
    \tau_i \frac{dg_i}{dt} = -g_i + \sum_{j,k}w_i\delta(t - t_j^k)\,,\\
    s(V) = \frac{1}{1 + ([\text{Mg}^{2+}]/\gamma)\exp{(-\beta V)}}\,.
\end{eqnarray}
Where $i$ can be $A$, $N$, or $I$, representing the AMPA, NMDA, and inhibitory channels, respectively. We also set $w_A = w_N = w_E$.  $s(V)$ is the sigmoidal modulating function, $[\text{Mg}^{2+}]$ is the concentration of magnesium ions, and $\gamma$ and $\beta$ are fitting parameters. For $\alpha$ to represent the proportion of AMPA and NMDA channels, we kept the number of their inputs equal, i.e., $K_A = K_N = K_E$. Like before the input rate $\nu_i$ is the same for all the population. Table \ref{tab:NMDA} displays the values of all the model parameters.

In the diffusion approximation and reduced to a one-dimensional Langevin equation, this model produces the following set of equations:
\begin{equation}
    \frac{dV}{dt} = \frac{(V - \mu(V))}{\tau(V)} + \sum_{i = A, N, I}h_i(V)\eta_i(t)\,,
\end{equation}
where
\begin{eqnarray*}
    &\tau(V) = \frac{\tau_L}{1 + (1 - \alpha)\mu_A + \alpha s(V)\mu_N + \mu_I}\,,\\
    &\mu(V) = \frac{\tau}{\tau_L}(E_L + (1 - \alpha)\mu_A E_E + \alpha s(V)\mu_N E_E + \mu_I E_I)\,,\\
    &h_A(V) = (1-\alpha)\frac{\sqrt{\tau_A}}{\tau_L}\sigma_A(E_E - V)\,,\\
    &h_N(V) = \alpha s(V)\frac{\sqrt{\tau_N}}{\tau_L}\sigma_N(E_E - V)\,,\\
    &h_I(V) = \frac{\sqrt{\tau_I}}{\tau_L}\sigma_I(E_I - V)\,.
\end{eqnarray*}

\begin{ruledtabular}\textbf{}
\begin{table}
    \centering
    \begin{tabular}{c c}
    \textbf{Parameter} & \textbf{Value} \\ \colrule 
    \textbf{Variable} & \textbf{Value} \\
    $\alpha$ & variable \\
    $E_L$ & -60mV  \\
    $E_E$ & 0mV  \\\
    $E_I$ & -80mV  \\
    $w_E$ & \{0.1, 0.5\}  \\
    $w_I$ & \{0.1, 0.4, 1.0, 10.0\}  \\
    $\tau_L$ & 20ms  \\
    $\tau_A$ & 1ms  \\
    $\tau_N$ & 100ms  \\
    $\tau_I$ & 10ms  \\
    $\tau_R$ & 2ms  \\
    $K_E$ & 400  \\
    $K_I$ & 100  \\
    $\theta$ & -50mV  \\
    $V_r$ & -60mV \\
    $\nu_i$ & \{5, 20, 50\}Hz\\
    $[\text{Mg}^{2+}]$ & 1mM \\
    $\gamma$ & 3.57mM \\
    $\beta$ & 0.062(mV)$^{-1}$
    \end{tabular}
    \caption{Table containing the set of parameters used for the NMDA model.}
    \label{tab:NMDA}
\end{table}
\end{ruledtabular}

\begin{figure*}
    \centering
    \includegraphics[width=0.99\textwidth]{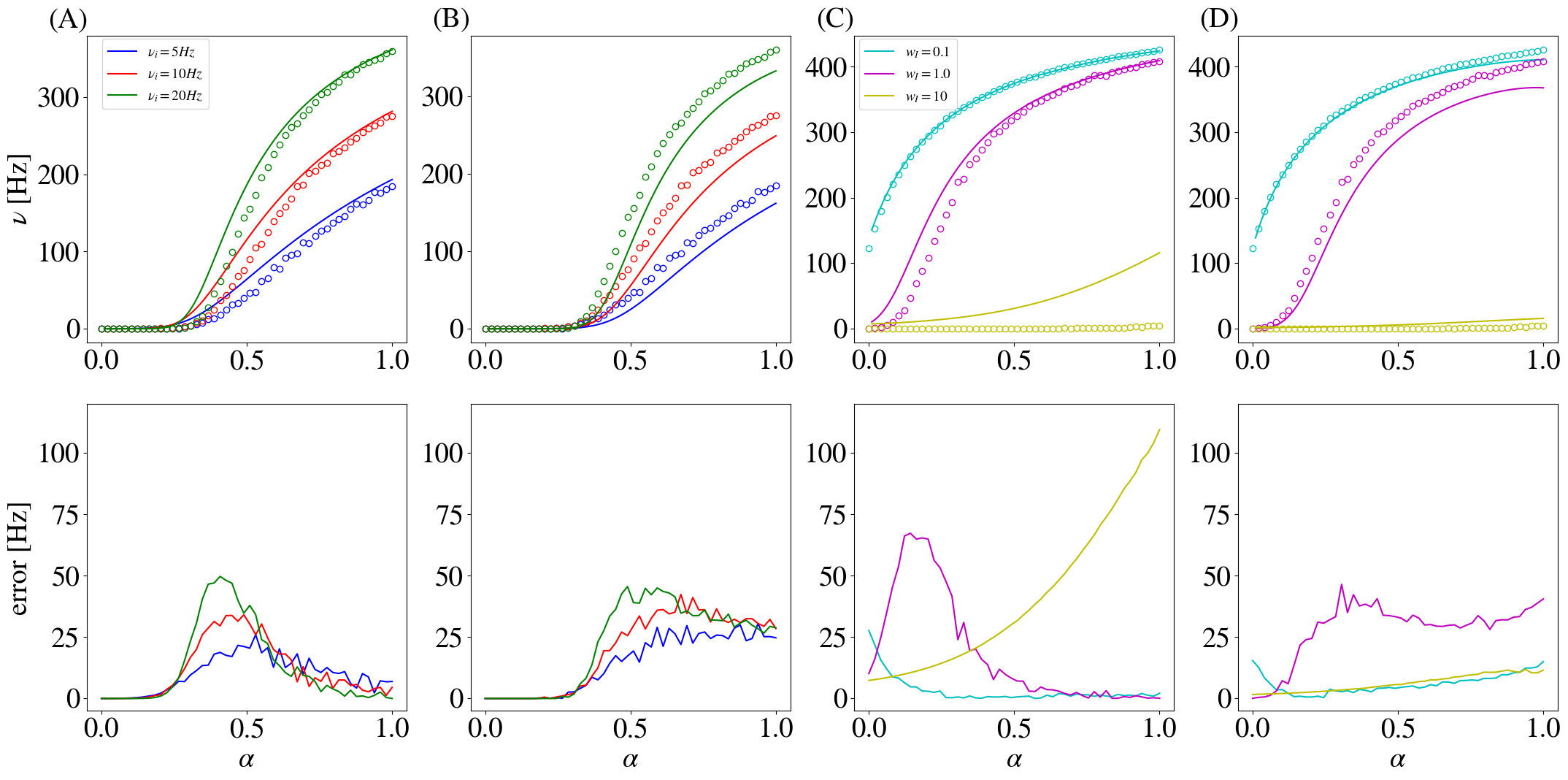}
    \caption{Comparison of the NMDA analytical model with simulations for three different values of input rate $\nu_I$ (columns (A) and (B)) and inhibitory synaptic weight $w_I$ (columns (C) and (D)) as a function of the interpolation parameter $\alpha$. In (A) and (B), we set $w_E=0.1$ and $w_I=0.4$, while in (C) and (D), we set $w_E=0.5$ and $\nu_i=5$Hz. Columns (A) and (C) assume the continuity of the distribution, while (B) and (D) use the double integration procedure. We calculate the error as the absolute distance between the simulation and analytical results.}
    \label{fig:NIF_firingRate}
\end{figure*}

Direct use of the effective time-constant approximation is no longer possible, since $h_N(V)$ is no longer linear, and the approximation loses its logic \cite{Richardson2005-it}. However, it is worth mentioning that for the NMDA model, it is possible to linearize the term $s(V)(V - E_E)$ around the average membrane potential, as was done by Brunel and Wang \cite{Brunel2001-ay}. Since we are trying to see how well the method performs for non-linear multiplicative noise, we will not perform the linearization and the approximation.

We start with the Fokker-Planck equation, 
\begin{equation}
    \frac{\partial P}{\partial t} = -\frac{\partial}{\partial V}\left[ W(V)P\right] + \sum_{i=A,N,I}\frac{\partial}{\partial V}h_i(V)\frac{\partial}{\partial V}(S_i(V)P) \,,
\end{equation}
where the functions $S_A(V)$, $S_N(V)$, and $S_I(V)$ are given by \eqref{eq:MN_S}.  $W(V)$ is no longer linear in $V$ and it results in different functional forms for the $S$ functions. The linear differential equation in $V$ is then
\begin{equation}
    \frac{\partial P_s}{\partial V} + B(V)P_s = -\nu H(V)\,,
\end{equation}
with coefficients
\begin{eqnarray}
    &B(V) = \frac{\sum_{i=A,N,I}h_i(V)S'_i(V) - W(V)}{\chi(V)}\,,\\
    &H(V) = \frac{\Theta(V - V_r)}{\chi(V)}\,,\\
    &\chi(V) = \sum_{i=A,N,I}h_i(V)S_i(V)\,.
\end{eqnarray}

\begin{figure*}
    \centering
    \includegraphics[width=0.99\textwidth]{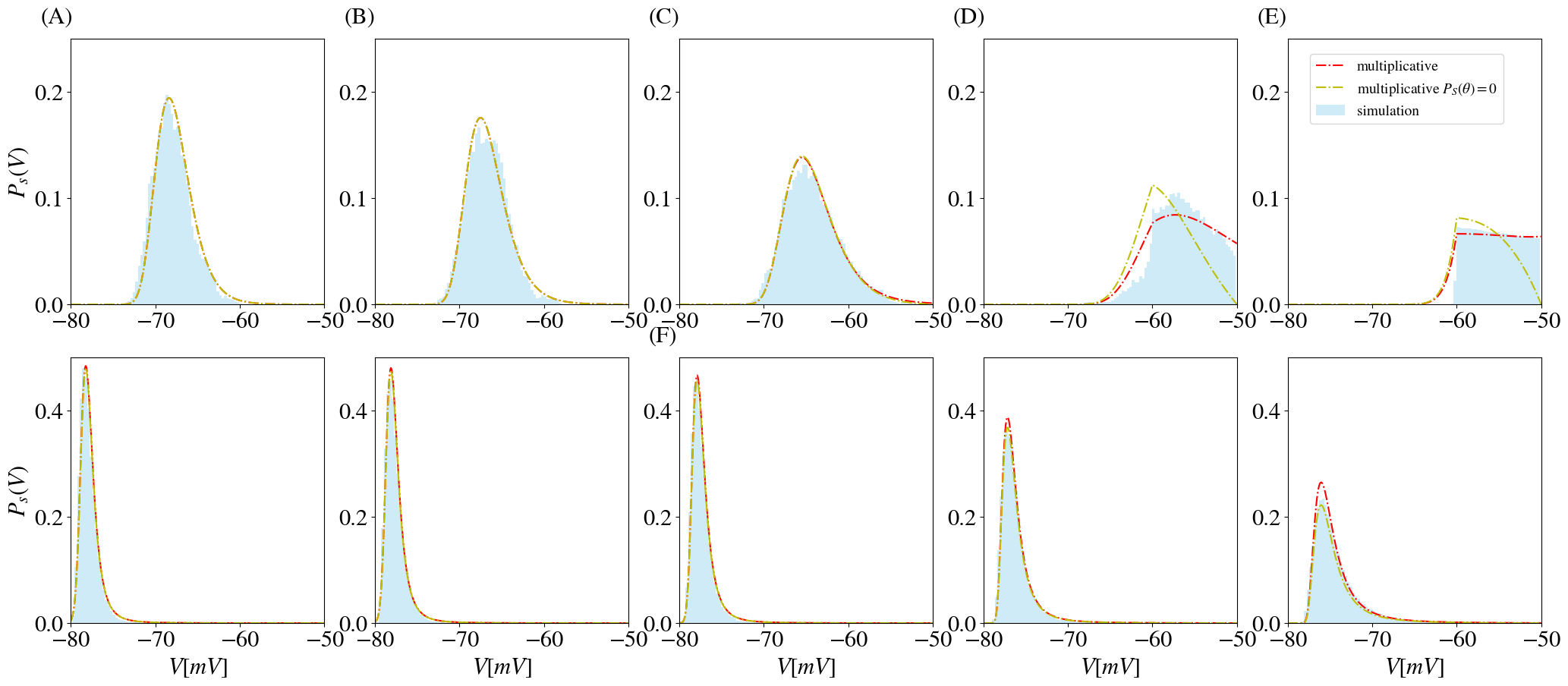}
    \caption{We present the stationary probability distribution for different values of the interpolation parameter $\alpha$: (A) $\alpha=0$, (B) $\alpha=0.1$, (C) $\alpha=0.2$, (D) $\alpha=0.5$, and (E) $\alpha=0.9$. The first row corresponds to the blue curves in Figure \ref{fig:NIF_firingRate}, with $\nu_i=5Hz$ and parameters from Table \ref{tab:NMDA}. The second row corresponds to the yellow curve on the same figure, with $w_I=10$, $w_E=5$, and $\nu_i=5Hz$. The legend indicates different assumptions in the model for each line.}
    \label{fig:NIF_dist_NMDA}
\end{figure*}

We proceed to test this model against simulation data for different input rates (figure \ref{fig:NIF_firingRate}A and \ref{fig:NIF_firingRate}B) and for different inhibitory weights (figure \ref{fig:NIF_firingRate}C and \ref{fig:NIF_firingRate}D). Given the highly nonlinear model, it is remarkable how good is the agreement between the mean-field results and the simulations for the majority of the cases. The transition region is where most of the error is concentrated. We can also observe an interesting behavior in this model. The use of the continuity assumption can result in smaller errors in some situations, especially when $\alpha \approx 1$. However, in the high inhibition regime,  the double integration procedure produces a more accurate result, with the calculated firing rate following close to the almost quiescent simulated neuron. Transitions are then slightly better estimated with the double integration procedure, which can be important for phase transition calculations. 

A crucial point has to be made regarding the condition for uniform convergence \eqref{FT:UCcondition} that is usually used as a metric for good behavior in approximation methods \cite{Fox1986-uq}. For the linear conductance-based models, it is obeyed for all values. However, the introduction of the NMDA nonlinearity makes this condition break for a large range of parameters as can be seen by numerical calculations of \eqref{FT:UCcondition}. Notably, the failure of the uniform convergence condition didn't affect the ability of the model to describe the system. But, for the method to work correctly it is necessary to deal with the divergence point at $1 - \tau_i\left( W'(y) - \frac{h_i'(y)}{h_i(y)}W(y)\right) = 0$. We were able to remove the divergence by excluding from the numeric integration an interval of $\pm 0.5$ around the value of $V$ that contains the divergence. No significant impact can be observed in the results when using this procedure, which indicates that the divergence cancels out in the integration.

Fig \ref{fig:NIF_dist_NMDA} displays the estimated probability distributions  for the NMDA model. The top row (corresponding to the blue curve in Figs. \ref{fig:NIF_firingRate}A and \ref{fig:NIF_firingRate}B), indicates that the estimation is good when the distribution don't interact heavily with the threshold but degrades when the interaction is significant. We see, nevertheless, that the continuity assumption causes a large error at the end of the left tale of the distribution, which doesn't occur as much with the double integration procedure. It still overestimates the probability of values slightly smaller than the reset potential. The bottom row (corresponding to the yellow curve in Figs. \ref{fig:NIF_firingRate}C and \ref{fig:NIF_firingRate}D) shows that the distribution remains mostly below the reset potential and little interaction with the threshold happens. The distribution spreads with higher $\alpha$ but the peak barely moves. The area of the red curve remains higher than the yellow, which helps explain the higher firing rate observed in \ref{fig:NIF_firingRate}C.

\section{Conclusion}
We developed a new method for constructing a transfer function for conductance-based integrate-and-fire neurons. This method is based on the mean-field Fokker-Planck approach and incorporates colored noise through the Fox Theory. We reduced the N-dimensional system into a single Langevin equation with colored and multiplicative noise and used the Fox Theory to construct an effective Fokker-Planck equation, which was solved to obtain stationary firing rates. We tested the method on two neuron models with progressive complexity: a standard conductance-based integrate-and-fire and a conductance-based integrate-and-fire with nonlinear NMDA channels.

For the standard conductance-based integrate-and-fire neuron, we compared mean-field results with firing rate data resulting from simulations as a function of the excitatory time constant. We found good agreement between the data and mean-field results in most scenarios, but the continuity assumption generated substantial errors in the transition regions. To correct that, we developed a double integration procedure that consists in estimating the probability density at the threshold with the first integration and using this value to start a backward integration to better estimate the distribution and the firing rate. We also observed that the effective time-constant approximation produced sharper transitions and the double integration procedure translated the curve to better match the transition point.

We then added a nonlinear NMDA channel to the model to test the method's effectiveness for nonlinear multiplicative terms. The method produced a good description of the simulated data even when outside the range of validity given by equation \eqref{FT:UCcondition}, but required extra care for some sets of parameters. Additionally, we discovered that in this particular scenario, the application of the double integration procedure does not always yield superior outcomes compared to utilizing the continuity assumption. While it generates a more accurate estimation in the transition region, it slightly underperforms for high $\alpha$ values.

When the membrane potential mean-field probability distributions are compared with simulation data, the effects of the continuity assumption become very clear. As expected, the distributions observed in the simulations are discontinuous. In fact, the discontinuity can occur not only at the threshold but also at the reset potential. Therefore, the continuous mean-field solution does not correctly estimate the distribution in most cases. The double integration procedure, however, produces better results, with $P(\theta)$ approximating the simulated one more accurately. We can also see that the effective time-constant approximation generates only small deviations from the full treatment, and is generally a good approximation.

From the analysis of the probability distribution, it is clear that the double integration procedure produces good estimations of the distribution at the threshold limit for most cases. However, when the neuron is in the mean-driven regime, the discontinuity at the reset potential is not taken into account. An appropriate study of the correct boundary conditions of the 1-D reduced Fokker-Planck equation can probably generate better results at the discontinuity points. Since the area of the distribution is related to the firing rate, this would probably improve the transfer function resulting from the method.

We have only tested our method on two types of neuron models, but it has the potential to be applied to a wider range of models. One possible extension, mentioned in the methods section, is the introduction of a nonlinear term in the drift expression. This nonlinear term can represent a quadratic \cite{Latham2000-mh} or an exponential integrate-and-fire neuron \cite{Fourcaud-Trocme2003-zv}, for example. It is also possible to introduce adaptation currents that depend on the spiking time of the modeled neuron. This would introduce the firing rate in the resulting Langevin equation, making a self-consistent treatment required.

It is also possible to look for non-stationary solutions when the input changes over time. For example, if we introduce an oscillatory Poissonian input, it is possible to construct a Fokker-Planck equation with the Fox theory, as the noise terms still have the same form but are now modulated by the input firing rate. However, only the stationary solutions of Fox theory are valid for all the ranges of time constants \cite{Jung1987-md}. The non-stationary solutions are valid in the small $\tau_i$ limit, and careful consideration of this fact is necessary when putting the method into practice.


\bibliography{apssamp}

\end{document}